\documentclass[a4wide]{elsart}
\usepackage{psfig}
\usepackage{graphics}

\begin{document}

\begin{frontmatter}
\title{\large \bf Study of intermediate velocity products in the  
Ar+Ni collisions between 52 and 95 A.MeV.\thanksref{Ganil}}
\thanks[Ganil]{Experiment performed at Ganil}

\author[LPC]{T.~Lefort \thanksref{Lefort}},
\author[SACLAY]{D.~Dor\'e},
\author[LPC]{D.~Cussol},
\author[LPC]{Y.G.~Ma \thanksref{Ma}},
\author[LPC]{J.~P\'eter},
\author[SACLAY]{R.~Dayras},
\author[SUBATECH]{M.~Assenard},
\author[GANIL]{G.~Auger},
\author[IPNO]{Ch.O.~Bacri},
\author[LPC]{F.~Bocage},
\author[LPC]{R.~Bougault},
\author[LPC]{R.~Brou},
\author[SACLAY]{Ph.~Buchet},
\author[SACLAY]{J.L.~Charvet},
\author[GANIL]{A.~Chbihi},
\author[LPC]{J.~Colin},
\author[IPNL]{A.~Demeyer},
\author[LPC]{D.~Durand},
\author[SUBATECH]{P.~Eudes},
\author[IPNO]{J.D.~Frankland},
\author[IPNL]{E.~Galichet},
\author[LPC]{E.~Genouin-Duhamel},
\author[IPNL]{E.~Gerlic},
\author[SUBATECH]{M.~Germain},
\author[SUBATECH]{D.~Gourio \thanksref{Gourio}},
\author[IPNL]{D.~Guinet},
\author[LPC]{B.~Hurst},
\author[IPNL]{P.~Lautesse},
\author[SUBATECH]{J.L.~Laville},
\author[LPC]{J.F.~Lecolley},
\author[GANIL]{A.~Le F\`evre},
\author[SACLAY]{R.~Legrain},
\author[LPC]{N.~Le Neindre},
\author[LPC]{O.~Lopez},
\author[LPC]{M.~Louvel},
\author[IPNL]{A.M.~Maskay},
\author[SACLAY]{L.~Nalpas},
\author[LPC]{A.D.~N'Guyen},
\author[IFIN]{M.~Parlog},
\author[IPNO]{E.~Plagnol},
\author[GANIL]{G.~Politi},
\author[SUBATECH]{A.~Rahmani},
\author[SUBATECH]{T.~Reposeur},
\author[NAP]{E.~Rosato},
\author[GANIL]{F.~Saint-Laurent \thanksref{Saint-Laurent}},
\author[GANIL]{S.~Salou},
\author[LPC]{J.C.~Steckmeyer},
\author[IPNL]{M.~Stern},
\author[IFIN]{G.~Tabacaru},
\author[LPC]{B.~Tamain},
\author[IPNO]{L.~Tassan-Got},
\author[GANIL]{O.~Tirel},
\author[LPC]{E.~Vient},
\author[SACLAY]{C.~Volant},
\author[GANIL]{J.P.~Wieleczko} and
\author[LPC]{A.~Wieloch \thanksref{Wieloch}},
\address[LPC]{LPC Caen
(IN2P3-CNRS/ISMRA et Universit\'e),
14050 Caen Cedex , France}
\address[SACLAY]{DAPNIA-SPhN, CEA/Saclay,
91191 Gif-sur-Yvette Cedex, France}
\address[SUBATECH]{SUBATECH (IN2P3-CNRS/Universit\'e),
44070 Nantes Cedex, France }
\address[GANIL]{GANIL (DSM-CEA/IN2P3-CNRS),
B.P. 5027, 14076 Caen Cedex 5, France}
\address[IPNO]{IPN Orsay (IN2P3-CNRS),
91406 Orsay Cedex, France}
\address[IPNL]{IPN Lyon (IN2P3-CNRS/Universit\'e),
69622 Villeurbanne Cedex, France}
\address[IFIN]{Nuclear Institute for Physics and Nuclear Engineering, 
Bucharest, Romania   }
\address[NAP]{Dipartimento di Scienze Fisiche, Univ. di Napoli,180126 Napoli, 
Italy.  }
\thanks[Lefort]{present address : Indiana University
    Cyclotron Facility, Bloomington, IN 47408, USA}
\thanks[Ma]{present address : Institute of Nuclear Research,
Shangai, China}
\thanks[Gourio]{present address : GSI, Postfach 110552, 64220 Darmstadt, 
Germany}
\thanks[Saint-Laurent]{present address : CEA, DRFC/STEP, CE Cadarache, 
13108 Saint-Paul-lez-Durance, France}
\thanks[Wieloch]{on leave of absence from Institute of Physics,
Jagiellonian University, Reymonta 4, 30059 Krak\'{o}w, Poland.}
\newpage
\begin{abstract}

 Intermediate velocity products in Ar+Ni collisions from 52 to 95  A.MeV 
are studied in an experiment performed at  the GANIL facility with 
the 4$\pi$ 
multidetector INDRA. It is shown that these emissions cannot be explained by 
statistical decays of the quasi-projectile and the quasi-target 
 in complete equilibrium. Three 
methods are used to isolate and characterize  intermediate velocity 
products. The  total mass 
 of these products increases with the violence of the collision 
and reaches  a large
fraction of the system mass
in mid-central collisions.  This mass is found 
independent of the incident energy, but strongly dependent  on 
the geometry of the collision. 
Finally it is shown that the kinematical
characteristics of  intermediate velocity products are weakly dependent 
 on the
experimental impact parameter,
but strongly dependent  on the incident energy. 
The observed trends are consistent with a participant-spectator like scenario 
or with neck emissions and/or break-up.



\end{abstract}

\end{frontmatter}

\section{\label{intro}\textbf{Introduction}}

Intermediate velocity products have shown to provide 
informations on
the properties of nuclear matter. The study of in-plane flow (also called
sideward flow) allows to determine the parameters of the nucleon-nucleon
interaction by comparing the theoretical calculations to the experimental data
\cite{Bert87,Buta95,Dan88,Doss86,Gus84,Gut90,He96,Krof92,Li96,DLM92,Ogil89,Pak97,Popes94,Shen93,Stock80,Sul90,West93,Rei97,Xu91,Zhang95,Ang97}.
In these studies, the in-plane flow is assumed to result from the 
particles emitted at
the first moments of the collision (also called ``mid-rapidity'' emissions or
``pre-equilibrium'' emissions). But these studies have also shown that the
flow parameter value is depending on the nature of the particle 
\cite{Li96,Pak97,West93}. 
To understand
this effect, a study of the emission mechanisms at mid-rapidity is needed.
Indeed, theoretical calculations \cite{Ono95} 
have shown that an interplay between the
mid-rapidity emissions and the particles emitted by the statistical decay of a
quasi-projectile and a quasi-target could explain the dependence of observed
in-plane flows on the particle nature. Other studies have shown that
the amount of matter emitted at mid-rapidity is also sensitive to the
parameters of the nucleon-nucleon interaction \cite{Gal98}. 
The aims of this article are: 
i) to understand the reaction mechanism(s) leading to the
intermediate velocity emissions, ii) to estimate the amount of matter emitted at
mid-rapidity to put constraints on theoretical calculations.
In addition, it is 
necessary to recognize and exclude the contribution of dynamical 
emissions in order to perform valid statistical analysis of hot nuclei 
decay modes as for caloric curve studies 
\cite{Ma97,Nat95,Mor96,Haug96,Kwia98} or binomial analysis \cite{Mor97}. 
These points 
underline the importance of studying intermediate velocity products.

If the overall  characteristics of reaction mechanisms are now well 
known at low and relativistic energies, they are still highly discussed and not
fully understood at intermediate energies. 
Below 20 A.MeV, pure binary processes, 
Deep Inelastic Collisions (DIC), are mainly observed for heavy systems while 
fusion process occurs for the light and medium mass systems in central and 
semi-central collisions \cite{Lef78}. Above 200 A.MeV, two excited 
spectators and a third source,  the so-called participant zone, 
are present in the final stage over almost the whole range of impact 
parameters \cite{Gus84,Sch74,Wes76}.  
   
In the intermediate energy range, experimental results \cite{Fuc94,Pet95} 
and theoretical calculations \cite{Pei94,Sob94,Col95_1,Col95_2} 
have shown that the reaction mechanism 
is mainly binary. The fusion cross section decreases with  increasing beam 
energy and represents at most a few percent of the reaction cross section at 
incident energies above 50 A.MeV \cite{Pet95,Beau96}. For a very broad range 
of impact parameters, the final products can be attributed to two excited 
outgoing nuclei, namely the quasi-projectile (QP) and the quasi-target 
(QT), accompanied by dynamical emissions localized between the QP and QT 
rapidities, i.e. around mid-rapidity
\cite{Pet95,Lanz98,Day89,Sauv94,Tsang86,Pet 2) 90}. 
This article will be devoted to the study of  these intermediate velocity
products.  We use this name to avoid confusion with ``mid-rapidity''
emission which is often associated with very fast emissions.

Several  scenarii were proposed to explain the origins of 
 intermediate velocity products. We are going to describe them as 
a function of the reaction time. 

In the first step of the collision, when the projectile and the target 
begin to overlap, some of their nucleons suffer nucleon-nucleon 
collisions in the interacting zone \cite{Fuc94}.      
Part of these nucleons get (or keep) a sufficiently high energy to escape 
from the attractive potential of the bulk  and are emitted around 
mid-rapidity. This emission was called ``prompt'' emission or preequilibrium 
emission and concerns mostly nucleons and light clusters ($\rm{^2H}$, 
$\rm{^3H}$, $\rm{^3He}$, $\rm{^4He}$, $\rm{^5Li}$ ...).  Cluster 
emission could be due to collisions of nucleons and/or clusters 
on pre-existing clusters. It could also be 
explained by the coalescence of prompt nucleons as predicted 
by AMD calculations \cite{Ono95}. 
            
After this stage, other  intermediate products  may come from the 
de-excitation of the overlap region.  Its decay 
can occur during the sticking of QP and QT \cite{Eud97} 
and/or after their separation. Two  scenarii are possible to explain 
the separation mechanism between the overlap region and the two partners: 
either the formation of a neck of matter between QP and QT, or 
a sharp and geometrical break-up between the QP and QT and the 
overlap region (i.e the participant-spectator scenario). The competition 
between these two  mechanisms is governed by the interaction time. 

If the di-nuclear system has enough time to reach a deformed shape, a 
neck of matter can be formed between the two interacting nuclei. Then, 
the rupture and/or the emission from the neck leads to the 
formation of products around mid-rapidity as observed 
\cite{Stu92,Lec95,Lar99,Dem96,Tok95,Che96,Mon94,Luk97} and 
predicted \cite{Sob94,Col95_1,Col95_2,Luk97} from peripheral to semi-central 
collisions below 50 A.MeV.
The shearing off may occur later on from one of the two outgoing nuclei.
In this case,  QP and/or QT are deformed along the axis which 
connects them and their decays are anisotropic:
one observes a preferential emission in the backward region of the 
QP frame and/or in the forward zone of the QT frame. This emission  
enhances the product yields around mid-rapidity. This process 
was observed from peripheral to semi-central collisions 
\cite{Luk97,Cas93,Ste95}
and was called fast emission or dynamic fission of the two outgoing 
nuclei. Obviously, one can observe in the 
same events both processes: dynamic fission and neck rupture.

If the system has not enough time to deform itself, then the reaction 
mechanism corresponds to  the 
participant-spectator scenario \cite{Sch74}. One observes 
three sources: two spectators and a participant zone. 
The participant zone comes from the stopping of nuclear matter 
in the overlap region between the two colliding nuclei. 
Intermediate velocity products come from the decay of this participant 
zone created at mid-rapidity. The spectators correspond 
to the remaining matter of initial projectile and target which 
conserve a large part of their initial rapidities.

Above 50 A.MeV, the transition from the neck 
or dynamic fission processes to the participant-spectator 
scenario is expected.  
The properties of this transition 
would probably 
depend on the viscosity of hot (and compressed ?) nuclear matter in 
the interaction zone. It would give information on the viscosity of 
nuclear matter and on the in medium nucleon-nucleon cross-section.

This paper presents the results obtained in the study of $\rm{^{36}Ar+
^{58}Ni}$ reactions between 52 and 95 A.MeV using the INDRA detector.
Following a brief description of the experimental conditions 
(section \ref{Setup}),
the event selection and the impact parameter sorting will be presented 
(section \ref{sorting}). Section \ref{signat} will show the evidence for 
 intermediate velocity products. Section \ref{estimation} 
will describe three methods  developed for the characterisation
of  these products. Quantitative  estimates of this 
emission 
will be presented in section \ref{caract}.
The energy dissipated by intermediate velocity products
will be examined in section \ref{energy}. 
Conclusions will be drawn in section \ref{summar}.

\section{\label{Setup}\textbf{Experimental set-up}}
The experiment was performed at the GANIL facility with the INDRA 
detector. The $\rm{^{36}Ar}$ beam impinged on a  self supporting 
193 $\rm{\mu g/cm^2 ~^{58}Ni}$ 
target. Typical beam intensities were 3-4 $\rm{\times 10^7}$ pps. 
A minimal bias trigger was used: the events are registered when at least 
 three charged particle detectors  are fired, for beam energies 
smaller than 83  A.MeV, and at least four above. 

The INDRA detector can be schematically described as a set of 17 detection 
rings centered on the beam axis. The detection of charged products was 
provided in each ring with two or three detection layers: the most forward 
ring, $\rm{2^{\circ} \le \theta_{lab} \le 3^{\circ}}$, is made of phoswich 
detectors (plastic scintillators NE102 + NE115); between $\rm{3^{\circ}}$ 
and $\rm{45^{\circ}}$ 
eight rings are constituted by three detector layers: ionization chambers, 
silicon and CsI(Tl); beyond $\rm{45^{\circ}}$, usually the eight remaining 
rings are made of double layers: ionization chambers and CsI(Tl) but for this
experiment the ionization chambers were not yet installed  for 
$\theta_{lab}$
above $\rm{90^{\circ}}$. The total number of 
detection cells is 336 and the overall geometrical 
efficiency of INDRA detector corresponds to 90\% of 4$\rm{\pi}$. A complete 
technical description of the INDRA detector and of its electronics  is 
given in \cite{Pou95}. Isotopic separation was achieved up to Z=4 in the 
last layer (CsI(Tl)). 
Charge identification was  possible up 
to Z=54 in the forward region ($\rm{3^{\circ}} \le \theta_{lab} \le 
\rm{45^{\circ}}$) and up to Z=16 for 
$45^{\circ} \le \rm{\theta_{lab} \le 90^{\circ}}$. 
Due to the absence of backward ionization chambers in 
$\rm{^{36}Ar}$ + $\rm{^{58}Ni}$ experiments,
fragments with Z>4 could not be separated above $90^\circ$.
The energy resolution is equal to 5\%  on average for CsI(Tl) 
and ionization chambers and better than 2\% for Silicon.
The INDRA detector capacities allow one to carry out an event by event 
analysis and to determine reliable global variables related 
to the impact parameter.

\section{\label{sorting}\textbf{Preliminary analysis}}

\subsection{\label{selec}Selection of "well measured" events}
The first step in the analysis was to select events in which sufficient 
information was collected. This was achieved by requiring that 
for each selected event, $\sum_{i=1}^{i=M} Z_i V^i_{par}$, where 
$Z_i$, $V^i_{par}$ are respectively the charge and the parallel velocity 
of particle $i$ and $M$ the multiplicity, be larger than 70\% of the 
initial $Z_{proj} \times V_{proj}$ of the projectile \cite{Pet 1) 90}.
The number of events kept by this selection represents respectively 
60 to 50\% of the total registered events from 52 to 95 A.MeV. 
The major part of the eliminated events are characterized by a low 
multiplicity value, a small total detected charge and a low value of 
$\sum_{i=1}^{i=M} Z_i V^i_{par}$. These removed events correspond to 
some peripheral collisions in which the QT residue is not
detected, the QP is lost in the forward 2$^{\circ}$ hole and only few light 
particles are detected.

Since we want to study the  properties of intermediate velocity
products as a function of impact parameter, we checked that this selection 
conserves the whole impact parameter range of registered events. We note 
that the multiplicity and the total 
transverse energy distributions of 
selected events cover the whole range of the respective raw 
distributions (i.e without any selection). 
If one assumes that the total multiplicity and total transverse energy 
are  good measurements of the violence of the collision, then this result 
indicates that the whole impact parameter range of registered events 
is kept in the selected events. 

Furthermore, for a proper study of intermediate velocity products, it 
is necessary to remove the fusion events. Indeed fusion nuclei, if any, 
are localized near the center-of-mass velocity and therefore 
their decay products are emitted in the same range of velocity as the 
intermediate velocity products. 
We eliminate most of the fusion events 
with a condition on the flow angle $\rm{\theta_{el}}$. 
This angle is defined as
the angle of the main axis 
of the ellipsoid which represents the energy distribution of all detected
products. One $\rm{\theta_{el}}$ is calculated for each event. 
For fusion events the $cos\left(\rm{\theta_{el}}\right)$ distribution 
is flat. Thus we removed the flat part of the 
$cos\left(\rm{\theta_{el}}\right)$ distribution
calculated at each incident energy, which corresponds to 
$\rm{\theta_{el}}$ values above $\rm{60^{\circ}}$ as in \cite{Mar97}. 
This represents 1\% of the total cross 
section at 52 A.MeV and becomes negligible above 52 A.MeV.

\subsection{\label{sorting_b}Event sorting}
The next step was to sort the events as a function of the violence of the 
collision. Since no global variable is perfectly correlated to the 
impact parameter, we will use two weakly correlated 
global variables related to 
the violence of the collision. The first one is the total detected 
transverse energy:

\begin{equation}
 \rm{E^{tot}_{tr}=\sum_{i=1}^M E_l^i \times (sin(\rm{\theta_i}))^2}
 \label{eq_etrans}
\end{equation}

where $\rm{M}$ is the multiplicity of the event, $\rm{E_l^i}$, 
$\rm{\theta_i}$ are respectively the kinetic energy 
 and the polar angle of the particle $i$ in the laboratory 
frame.

The events were sorted in $\rm{E^{tot}_{tr}}$ bins. 
Assuming 
a monotonic relation
between $\rm{E^{tot}_{tr}}$ and the impact parameter, the 
cross section  associated to each bin can be expressed as an experimentally 
estimated impact parameter  \cite{Pet 1) 90}:

\begin{equation}
\rm{b_{exp}}\left(\rm{E^{tot}_{tr}}\right) =
\sqrt{
\frac{1}{\pi} \int_{\rm{E^{tot}_{tr}}}^{\rm{E^{tot}_{tr}} Max}
\sigma\left(\rm{e^{tot}_{tr}}\right) \quad d\rm{e^{tot}_{tr}} } 
\end{equation}

The first bin, associated to the highest $\rm{E^{tot}_{tr}}$, with a 
cross section of 31.4 mb, is linked to an experimental parameter 
$\rm{b_{exp}}$ $\leq$ 1 fm, the second bin is linked to 1 fm $\leq$ 
$\rm{b_{exp}}$ $\leq$ 2 fm and so on, up to the last bin which corresponds 
to $\rm{b_{exp}}$ $\geq$ 8 fm. This correspondence is established with the
 $\rm{E^{tot}_{tr}}$ distribution of all
detected events. The total cross section has been determined experimentally. 

 A second global variable has been used for the cases where the studied
observable is too strongly correlated to $\rm{E^{tot}_{tr}}$, as in section
\ref{energy}. 
It is an energy per nucleon 
calculated event by event in the center-of-mass 
of fragments with A $>$~3 emitted above mid-rapidity \cite{Ma97}: 
\begin{equation}
 {\mathcal E}_{\rm f}= \left\lbrack
   \sum_{i=1}^{\rm{M_f}} E_c^i + C
   \right\rbrack / \sum_{i=1}^{\rm{M_f}} A_i  
 \label{eq_Ef}
\end{equation}
where $\rm{M_f}$ is the multiplicity of particles emitted above the
mid-rapidity, $\rm{E_c^i}$, 
the kinetic energy in the fragments  (A $>$~3) frame. 
 The quantity C 
includes the mass balance
of emitted products and the estimated kinetic energy of neutrons 
assumed to be equal to the kinetic energy of protons minus the Coulomb
energy. 

$\sum_{i=1}^{\rm{M_f}} A_i$ is close to $A_{QP}$ and ${\mathcal E}_{\rm f}$ is 
an estimation of the QP excitation energy per nucleon if the intermediate 
velocity products contribution and the overlap of the QP and QT 
emissions are quantitatively weak.
As we use this variable  only to 
estimate the violence of the collision, we will not discuss the 
 meaning  of the obtained values  of ${\mathcal E}_{\rm f}$.  
The highest ${\mathcal E}_{\rm f}$ are certainly
reached in violent central collisions and the lowest values in gentle 
peripheral collisions. However the correlation between ${\mathcal E}_{\rm f}$ 
and the impact parameter is 
broader than with $\rm{E^{tot}_{tr}}$ since only one half of
the detected products are used to determine ${\mathcal E}_{\rm f}$, whereas all the
detected products are used to calculate $\rm{E^{tot}_{tr}}$.
Thus we will not  associate values 
of $\rm{b_{exp}}$ with this sorting, even if in average 
${\mathcal E}_{\rm f}$ is correlated to the impact parameter.

\section{\label{signat}\textbf{Evidence for  intermediate 
velocity products}}

\subsection{Kinematical properties of the emitted products}

\begin{figure}[h]
\begin{center}
\psfig{file=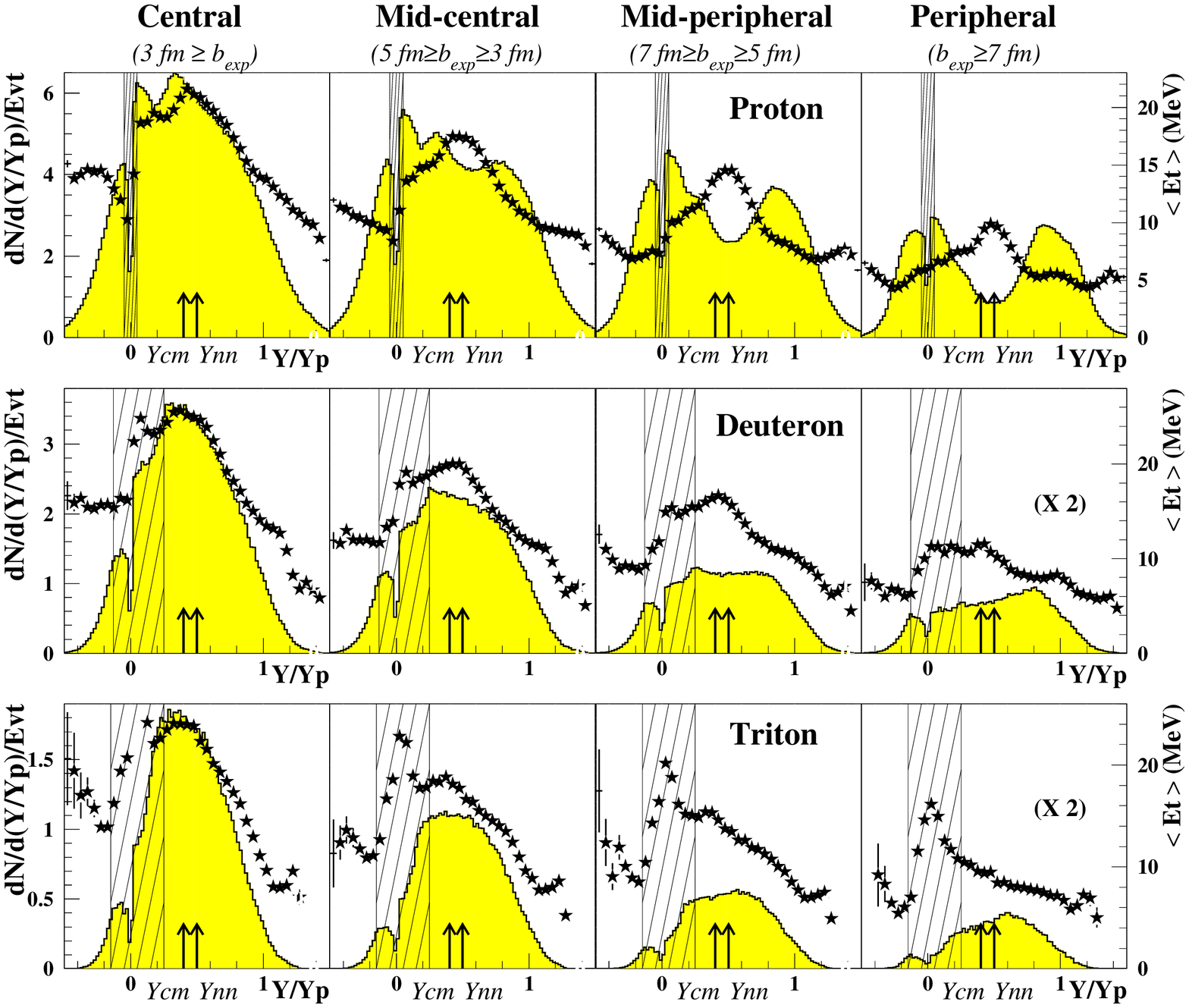,width=16.cm}
\caption[]{\label{rapet1}\rm{
 Differential multiplicities 
(grey histograms and left hand scale)
as a function of the reduced rapidity Y/Yp for p, d, t particles from 
central to peripheral collisions. Yp is the rapidity of the projectile.  
In this representation, the target, the
nucleon-nucleon center-of-mass, the system center-of-mass and the 
projectile have respectively reduced rapidities of
0, 0.385  (left arrow), 0.5 (right arrow) and 1.0. 
The stars represent the mean transverse energies 
(right hand scale) as a function of the reduced rapidity. 
 The hatched area around the target rapidity are affected by particle 
identification thresholds and the dead zone around the
target at $\theta$=$\rm{90^{\circ}}$.}}
\end{center}
\end{figure}

\begin{figure}[h]
\begin{center}
\psfig{file=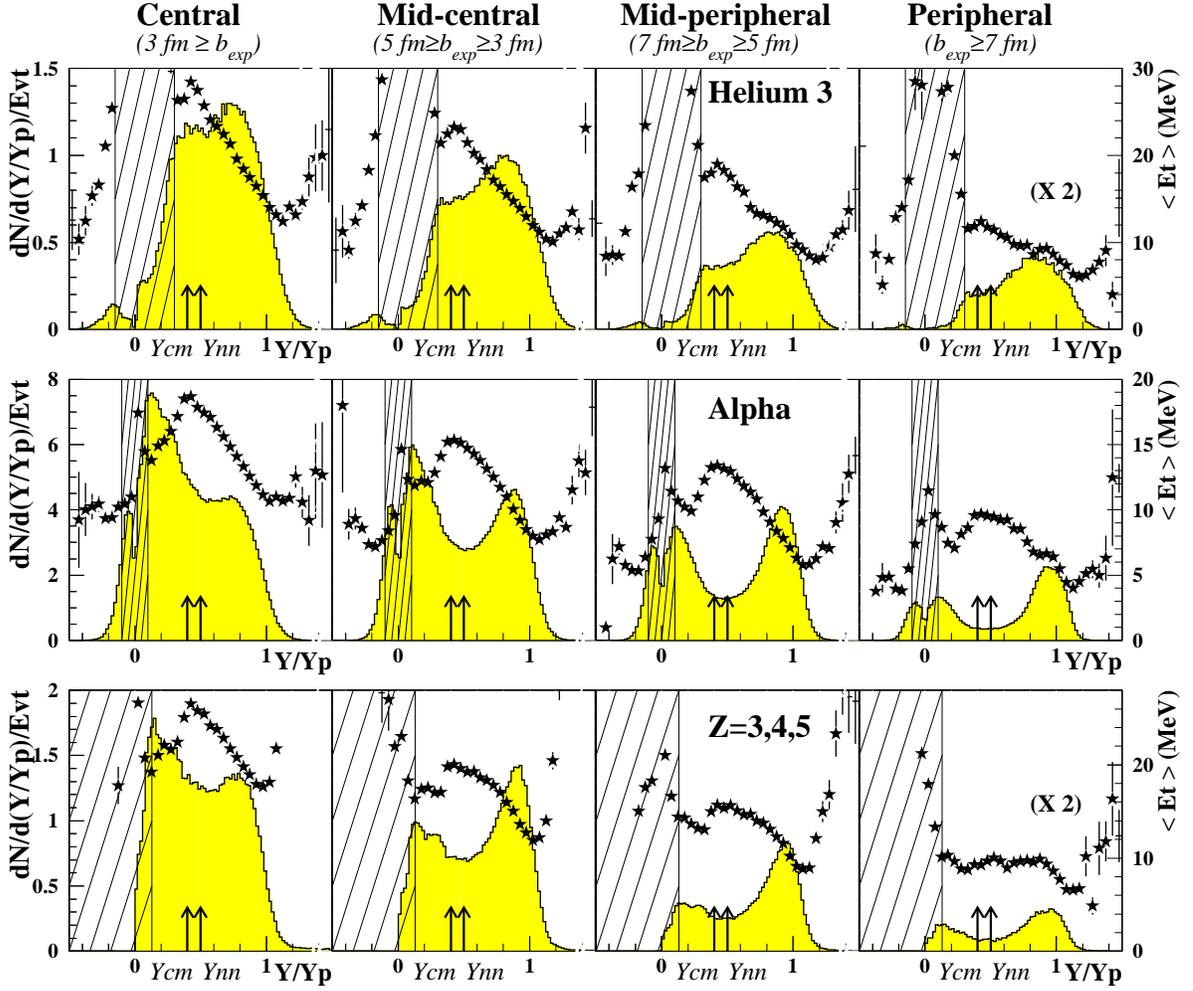,width=16cm}
\caption[]{\label{rapet2}\rm{Rapidity distributions in the laboratory frame 
and mean transverse energy of $\rm{^{3}He}$, $\rm{^{4}He}$ and Z=3,4,5 
determined at 74 A.MeV. See figure 1 for details.}}
\end{center}
\end{figure}

In order to have an overall view of production mechanisms 
in heavy-ion collisions, it is interesting to examine  the 
properties  of emitted products along and perpendicular 
to the beam axis. In figures
\ref{rapet1} and \ref{rapet2}  results are shown 
for four $\rm{b_{exp}(\rm{E^{tot}_{tr}})}$ bins. 
The grey histograms correspond to the reduced 
rapidity ($\rm{Y/Y_p}$,
where $\rm{Y_p}$ is the initial rapidity of the projectile)
distributions in the laboratory frame for 
protons (p), deuterons (d) , tritons (t), helium 3 ($\rm{^{3}He}$), alpha
particles ($\rm{^{4}He}$) and Z=3,4,5 at 74 A.MeV. The left vertical 
scale corresponds to the differential  multiplicity per event.
In this intermediate energy range, the rapidity value is slightly 
larger than the reduced velocity value ($\rm{\beta = \frac{v}{c}}$). 
In these figures are also shown 
the mean transverse energies $<$Et$>$~of each 
product versus $\rm{Y/Y_p}$ (black stars and right hand scale). 
Indeed, such a plot was shown to provide a clear and direct evidence 
for the presence of intermediate velocity products
different from products emitted by the equilibrated QP and QT \cite{Ang97}.
The hatched areas correspond to the reduced rapidity ranges affected by the
thresholds of the detectors.

As expected, one observes two hills in the reduced rapidity 
distributions (grey histograms on figures \ref{rapet1} 
and \ref{rapet2})   
of p, alpha and Z=3,4,5  
near the rapidities of the projectile and target. 
They correspond to the QP and QT de-excitation. One also notices the 
presence of products near mid-rapidity ($\rm{Y/Y_p}$=0.5). 
Most of p, alpha, Z=3,4,5 
come from the two partners de-excitation. On the other hand, 
d and t are mainly emitted around mid-rapidity. For $\rm{^{3}He}$, 
the situation is intermediate.

In addition, for every product we observe a maximum in $<$\rm{Et}$>$ 
around mid-rapidity (black stars on figures \ref{rapet1} 
and \ref{rapet2}). 
This bump indicates that light particles and Z=3,4,5 
emitted in the mid-rapidity region are more energetic than particles 
emitted by the two  main partners (QP and QT, as it was observed with 
the $\rm{^{36}Ar+^{27}Al}$ system at similar incident energies \cite{Ang97}). 
One also notes that the energy of particles at mid-rapidity
decreases when the estimated impact parameter 
increases. This decrease is due in part to the autocorrelation between the 
global variable (total transverse energy) used to sort the events and the 
transverse energy of the studied products. To avoid this problem, 
as already  mentioned in the
previous section, we will properly study the 
energetic properties of  intermediate velocity products
in section \ref{energy}
with the  ${\mathcal E}_{\rm f}$ sorting variable.
The large increase in $<$\rm{Et}$>$ observed near the target 
rapidity is related to the identification
thresholds which select high energy particles.
The increase of $<\rm{Et}>$ observed in the forward area of the
quasi-projectile and in the backward area of the quasi-target has been already
observed for the ${^{36}Ar+^{27}Al}$ system in \cite{Ang97}. It is not really
understood and could be an experimental indication of the presence of 
promptly
emitted particles in these rapidity areas \cite{Eud97}.  

The same kinematical trends have been observed from 52  
up to 95 A.MeV. 
In summary, these kinematical properties suggest the existence of an 
emission 
 at intermediate velocities (around mid-rapidity) which does not come 
from the statistical decay 
of the two partners (QP and QT). 

\subsection{Comparisons to a simulation}
 
In order to confirm  that the intermediate velocity products are not 
only evaporated by the QP and the QT, we have 
performed a simulation for the $\rm{^{36}Ar+^{58}Ni}$ reaction at 74 A.MeV.
Indeed, the observed kinematical properties could in principle be explained 
by a fast statistical emission from the QP and the QT: 

\begin{itemize}
\item{Firstly, if the products are emitted in the earlier stages of 
the collision, then the two partners are close to each 
other in space. Therefore the emitted products are 
accelerated by the Coulomb fields of the two partners. This Coulomb boost could 
possibly explain the bump observed in $<$\rm{Et}$>$ around $\rm{Y/Y_p}$=0.5.}
\item{Secondly, these products could be emitted by the QP and the QT 
before they reach their asymptotic velocities. 
Consequently, this fast statistical component 
could enhance the product yield between the final rapidities of QP and 
QT residues. This hypothesis is particularly valid for d,t and $\rm{^{3}He}$ 
which due to their higher separation energy than respectively p and alpha 
particles \cite{Bou97}, are essentially emitted in the first  decay 
steps of the hot nuclei decay. Thus d,t and $\rm{^3He}$ could be emitted 
near center-of-mass rapidity as it is observed.}
\end{itemize}

The calculations  assuming a pure binary scenario have been done with 
the statistical code SIMON \cite{Dur92} which has been sucessfully 
used in its standard version in \cite{Ma97,Siw98}. No dynamical emission 
has been included in the calculations. To compare the  results of the 
calculations with the experimental data, we used exactly 
the same procedure as in the experiment: the events  were "filtered" 
to simulate the effects of the INDRA detector and they  were selected and 
sorted as in the experimental data. Furthermore, to enhance 
the contribution of the statistical emissions at mid-rapidity
and to understand and enhance
the role played by Coulomb effects we have  divided the  statistical 
emission time of QP and QT by 100. 

\begin{figure}[h]
\begin{center}
\psfig{file=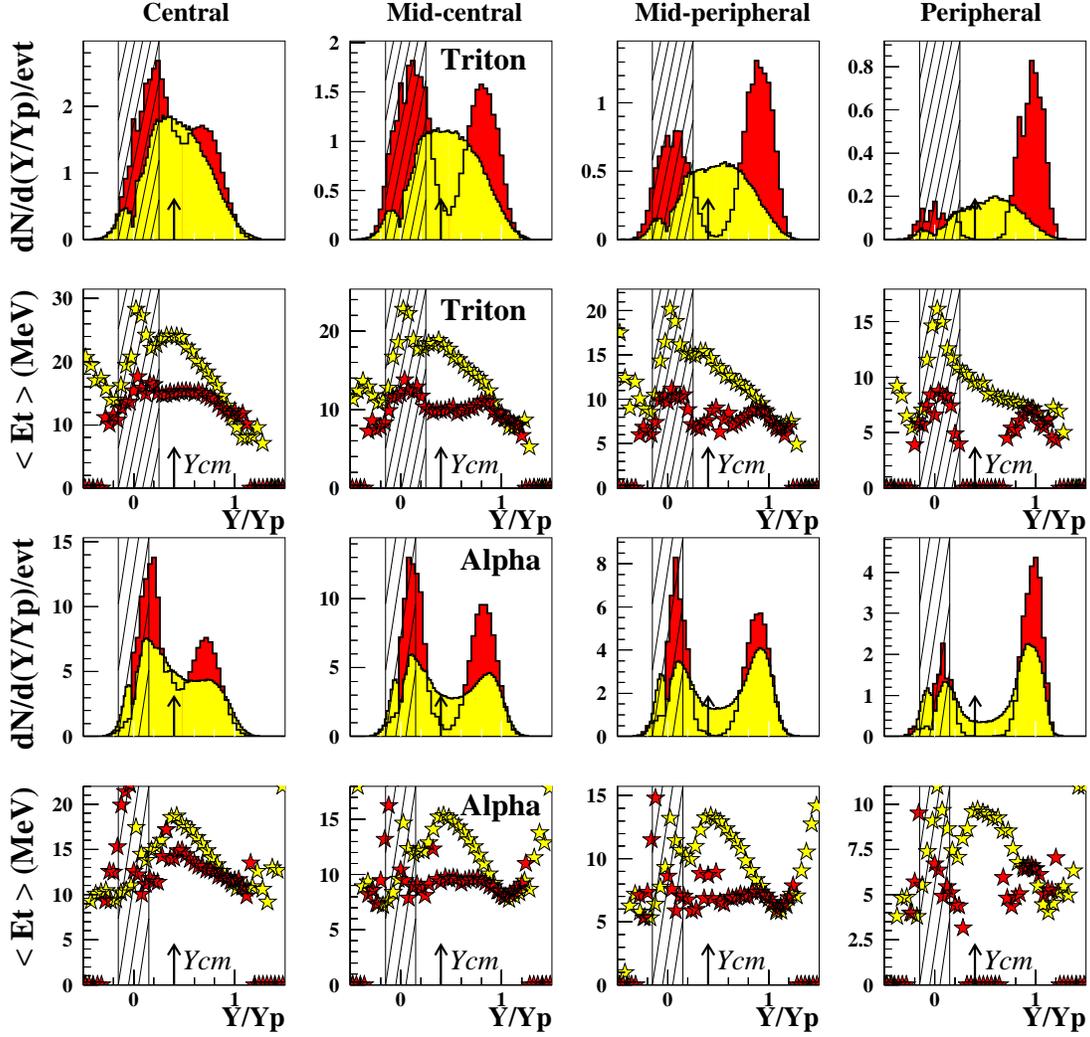,width=16cm}
\caption[]{\label{simrapet}\rm{Comparison between the experimental 
data and the simulation of purely binary collisions for the 
$\rm{^{36}Ar+^{58}Ni}$ reactions at 74 A.MeV. Rapidity distributions 
in the  laboratory frame (filled histograms) and mean transverse energy 
(stars) are plotted for tritons in the upper panel and for alpha particles 
in the lower panel. The experimental data are in light grey, the 
simulation in dark grey. The dashed zone corresponds to the rapidity 
domain where INDRA's  thresholds do not allow to  identify 
accurately the emitted products. The definition of the horizontal 
and vertical scales are identical to those of figures 1 and 2.}}
\end{center}
\end{figure}

The results are displayed in figure \ref{simrapet}, for two kinds
of products differing in  their experimental distributions: those 
which are 
mostly coming from the de-excitation of the 
two partners (alpha particles) and those which are essentially emitted 
around $\rm{Y/Y_p}$=0.5 (tritons). In both cases, we note that the number 
and the mean transverse energy of the emitted products 
 at intermediate velocities 
are strongly underestimated in the calculations.

In this simulation, the peaks near the QT and QP rapidities 
are found slightly  asymmetric: this is due to
particles which are emitted before the full acceleration of the QP and the QT.
  In the calculations, 
this fast statistical emission (when emission
times are divided by 100) represents less than 5\% 
 of the standard statistical emission from the QP.
   This asymmetry disappears with
standard statistical emission times. We notice 
that this asymmetry is more pronounced in experimental data  and then cannot
be explained by the statistical decay of the QP and the QT. 

In conclusion, in addition to the statistical decay of the two partners
there are in the $\rm{^{36}Ar+^{58}Ni}$ collisions additional emissions 
which lead to the formation of products around mid-rapidity. 
As observed in figure \ref{rapet1}, the detected d's and t's come mainly from 
the  intermediate velocity zone. So, the properties 
of these particles could reflect the fundamental properties of  
 intermediate velocity products in the $\rm{^{36}Ar+^{58}Ni}$ collisions.

For the sake of simplicity, we now define two kinds 
of emission. The first one is attributed to the statistical decay 
of the two partners after complete equilibrium of all degrees 
of freedom including the slow ones (shape). This component has been found 
 to be symmetrical in the QP or the QT frame. We call it 
 statistical emissions  (labelled SE in the figures). 
The second one is attributed to other 
emissions which  differ from the statistical decay of QP and QT. 
We call this component  intermediate velocity products (labelled IVP 
in the figures). 
One should keep in mind  the possible occurrence of several production 
processes, as described in the introduction.  There will be no attempt to
distinguish the specific contribution of each possible process.

\section{\label{estimation}\textbf{Separation of  intermediate velocity 
products and  statistical emission components}}
In order to quantify the  contribution of the intermediate velocity products, 
we have used three methods.
Two  of them are  based on the shape of the parallel 
rapidity distributions of 
detected products. The third one is 
based on a three source fit of the kinetic energy spectra.
Because of the thresholds which do not allow one to detect the slowest 
products (and therefore affect the shape of rapidity distributions) and 
in order to avoid the contribution of the evaporated particles from 
the QT, only emitted products with a reduced rapidity  ($Y_r 
= Y / Y_{proj}$) greater than 0.5 
will be used for the two methods based on rapidity distribution
shapes. The whole range of reduced rapidity will be used for the three
source fit method,  but only the values corresponding to the particles
emitted above $Y_r$=0.5 will be presented.

\subsection{Subtraction of the statistical decay of the QP: method  EVAP}
\begin{figure}[h]
\begin{center}
\psfig{file=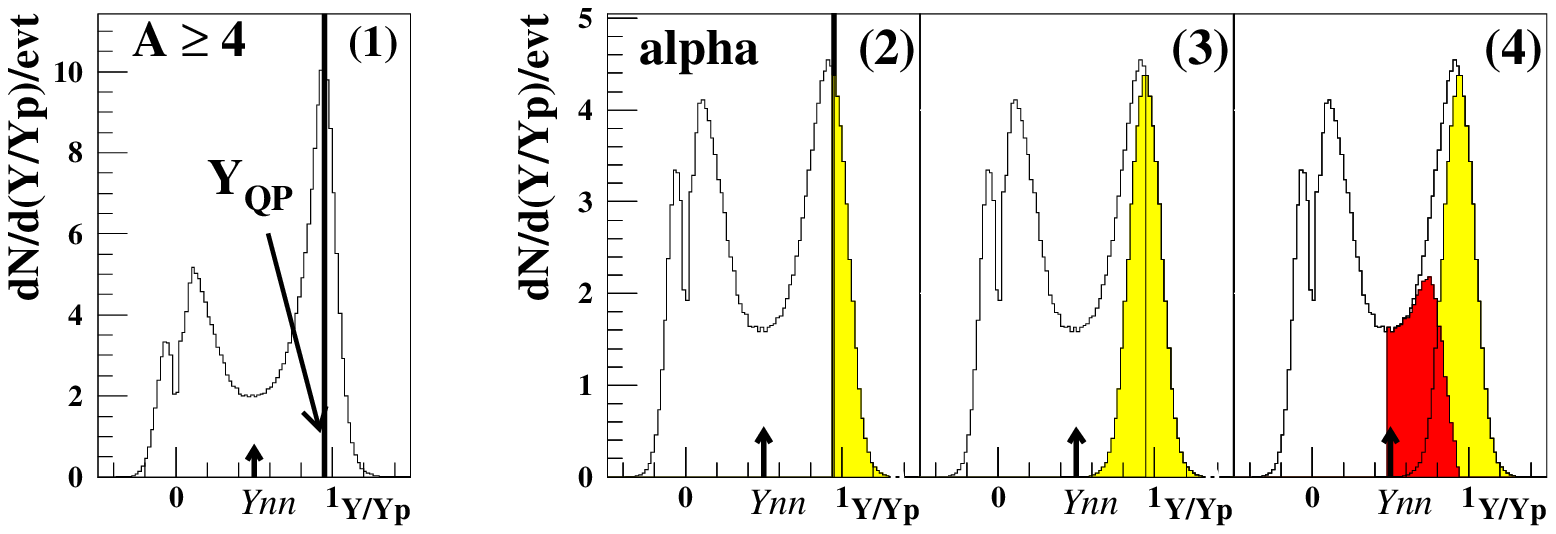,width=15cm}
\caption[]{\label{metE}\rm{Successive steps of method  EVAP 
for the determination
of the  intermediate velocity products component. 
This figure was established at 74 A.MeV for 
mid-peripheral collisions. See text for details.}}
\end{center}
\end{figure}

The first method, labelled  EVAP, consists in removing the 
statistical emission component 
from the QP. The successive steps of method  EVAP are displayed in figure
\ref{metE}.  

In the first stage, we determine the most probable QP's rapidity. 
Owing to an important  intermediate velocity products 
component for A $\rm{\le}$ 3 (see figures \ref{rapet1} and \ref{rapet2}) 
,we assume that the maximum observed on the A $\rm{\ge}$ 4 rapidity 
distributions above mid-rapidity corresponds to the QP rapidity (step 1). 
Then, we assume that all 
products which are emitted with a rapidity greater than the QP's 
rapidity come from the statistical decay of the QP (step 2). 
The backward evaporated component of the QP is 
obtained by symmetrizing the forward part relative to the QP rapidity. 
The total  statistical emission component is the sum of the backward and 
the forward component (step 3). Finally, to determine the intermediate 
velocity component we subtract the evaporated one from the total 
rapidity distributions (step 4).

One observes an excess of particles emitted below and close to 
the QP rapidity (shown in step 4 of figure \ref{metE}). In order to 
check that this excess is not due to a pollution of more dissipative 
events (with a lower QP rapidity) which would have been badly sorted,
we have selected events with fragments at a rapidity just below 
and close to the most probable QP rapidity. This selection has been done 
for each impact parameter value. If the selected fragments correspond to 
some more damped QP then 
the rapidity distributions of alpha particles 
emitted in coincidence should be centered around the rapidity of  
these fragments. It turned out that the alpha particles are emitted around 
the most probable QP rapidity. Thus the excess of particles emitted 
below and close to the most probable QP rapidity cannot be explained 
by the decay of more damped QP.

Some of the dynamically emitted particles  may be localized around the QP 
rapidity zone. Therefore, with this method, the number of evaporated 
particles in the QP forward hemisphere may be overestimated and so the 
total evaporated component. We thus obtained a lower limit of 
intermediate velocity products contribution with this method.       

\subsection{Direct estimation of  the intermediate velocity component: 
method DIRECT}
\begin{figure}[h]
\begin{center}
\psfig{file=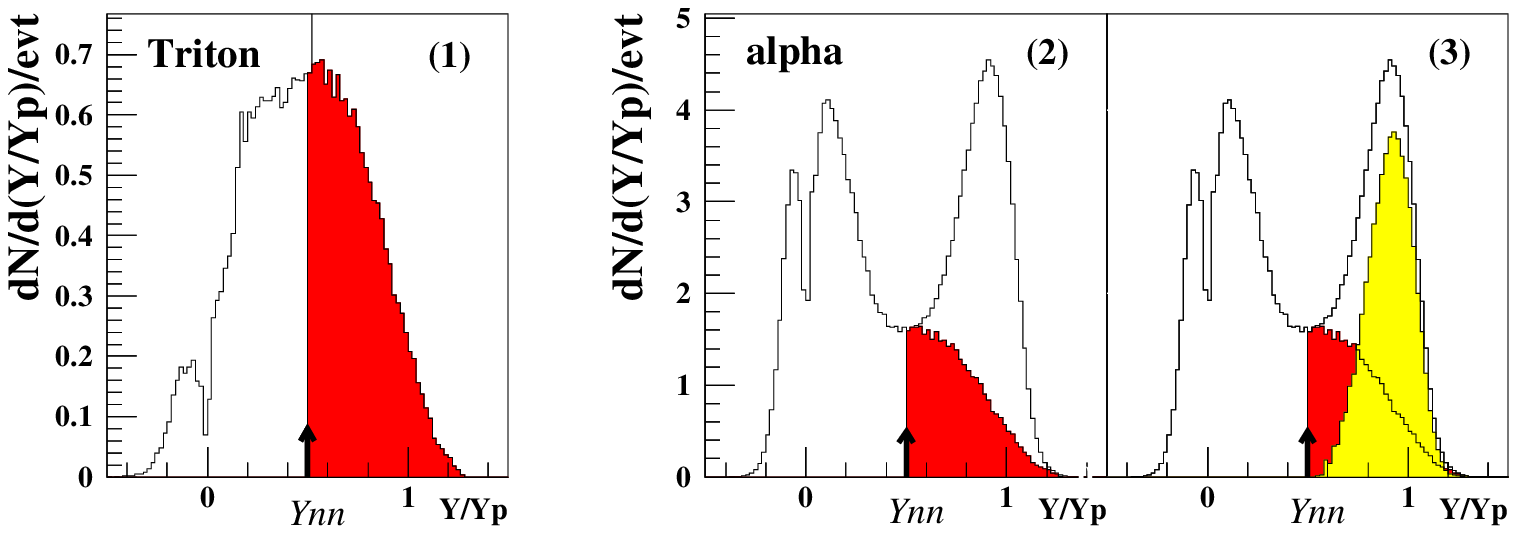,width=15cm}
\caption[]{\label{metM}\rm{Succesive steps of method  DIRECT for 
the determination
of the  intermediate velocity products component. 
This figure was established at 74 A.MeV for 
mid-peripheral collisions. See text for details.}}
\end{center}
\end{figure}

The second method, labelled DIRECT, consists in extracting directly 
the intermediate velocity component. 
The successive steps of method DIRECT 
are displayed in figure \ref{metM}.  

Figure \ref{rapet1}  (lower panel)  shows that the triton rapidity 
distributions  are  roughly 
symmetric around mid-rapidity. So, two assumptions are made: firstly, all 
tritons  are intermediate velocity products  
(step 1), secondly, the  shapes 
of the rapidity 
distributions of all  intermediate rapidity products 
are  homothetic to those of the tritons. To determine
the  intermediate velocity component of 
the other products, we  normalize at mid-rapidity the 
triton's rapidity distribution shape on their rapidity distributions
(step 2). Then we subtract this  intermediate velocity 
component from the total 
rapidity distribution to determine the evaporated
component (step 3).     
 
Some of the emitted tritons come from the statistical de-excitation of the QP
(statistical emissions component). 
Therefore, the  intermediate velocity component of the 
 tritons and 
of the other products are overestimated, thus we get an upper limit of 
 the intermediate velocity component.   

Since methods  EVAP and  DIRECT are based on independent 
 assumptions, we can check 
the  consistency between the two methods.
The method  DIRECT is based on two  hypotheses. The first one (step 1) was 
confirmed by the results obtained with method  EVAP: more than 70 \% of 
the emitted tritons  come from the  intermediate velocity 
component (except at 
52 A.MeV in the three bins of highest total transverse energy i.e 
$\rm{b_{exp}}$=1,2,3 fm). 
The second one (step 2) was also confirmed by method  EVAP for p, d 
and t: the rapidity distribution shape of their 
 intermediate velocity component, 
determined with method  EVAP, is identical to the triton rapidity 
distribution shape. On the other hand, for alpha particles and the 
Z=3,4,5, we have observed an enhancement just below the QP's rapidity. 
It may correspond to the dynamical break-up of the QP.  This 
contribution, which is not totally taken into account with method  DIRECT, 
is not large.

\subsection{The three source fit method:}

Three sources fit methods  (labelled TSF) have been extensively used in the past
\cite{Milk93,Wile91,Jac87} to obtain contributions, temperatures, velocities, 
Coulomb barriers of different sources of emission. The good granularity,
energy resolution and isotopic separation provided by INDRA should
allow a better adjustment of the fit parameters. We assume that QP, QT
and  intermediate velocity components 
can be approximated by three thermalised sources. It is only a way to model
the different sources of emission since we do not know if thermal 
equilibrium is achieved  
 for intermediate velocity products. 
The sum of three  Maxwellian distributions is
used to fit the energy spectra of emitted
light particles. In the laboratory frame, it becomes :
\begin{equation}
 d^2\sigma/dEd\Omega = \sum_{i=1,3} N_{i} 
\sqrt{E_{l}} \hskip 0.2cm exp-((E_{l} + Es_{i} - 2 
\sqrt{(E_{l} Es_{i})} cos(\theta _{l}))/T_{i}) 
\label{eq_tsf}
\end{equation}
N$_{i}$, Es$_{i}$ (=1/2 M$_{part}$V$_{source}^2$) and 
T$_{i}$ being adjustable parameters and E$_{l}$ and $\theta$$_{l}$,  
the energy and angle of the 
emitted particle in the laboratory.
The 9 parameters can be reduced to 6
assuming that :  i) QP and QT have the same temperature \cite{Lanz98}, 
 ii) QP and
QT parallel velocities are linked 
(V$_{QT}$=(M$_{proj}$/M$_{target}$)(V$_{proj}$-V$_{QP}$)),
 iii) two N$_{i}$ can be expressed relatively to the third one.

\begin{figure}[h]
\begin{center}
\psfig{file=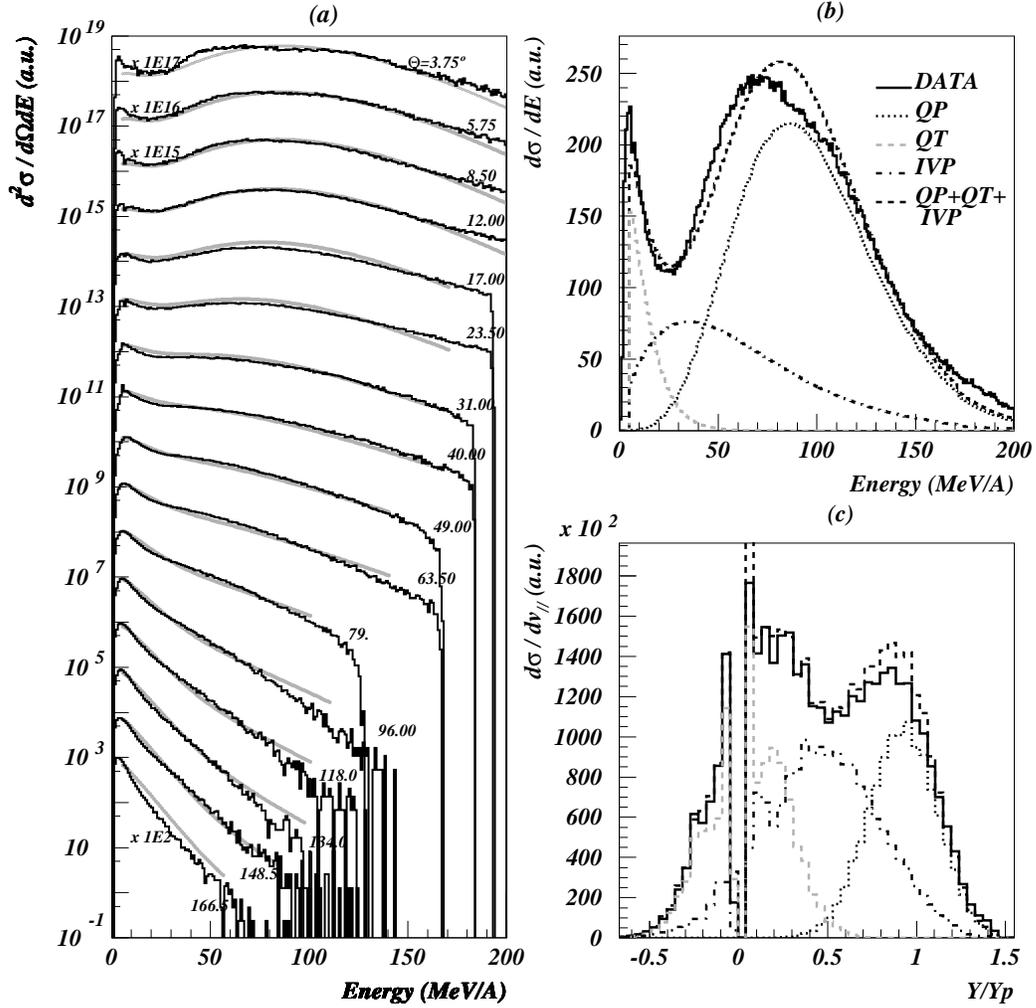,width=15.cm}
\end{center}
\vskip -1.cm
\caption[] {\label{f:tsf} { Protons at $b_{exp}$=5 fm, for $\rm{^{36}Ar + ^{58}Ni}$ 
collisions at 95 A.MeV. 
(a) Energy  spectra at different angles.
The histograms correspond to the experimental data and  the dotted lines to
the results of the  fits.
(b) Total proton energy spectrum in the laboratory frame. 
(c) Parallel velocity distribution. The curves are the results of the
three source fits.}}

\end{figure}

For each light particle  species and each impact parameter bin, 
we fit the sixteen energy
spectra (rings 2 to 17) with relation (\ref{eq_tsf}).
 As the Coulomb barriers have been neglected, 
for large impact parameters (${\rm b_{exp} \ge 8 fm}$  at 95 A.MeV and 
${\rm b_{exp} \ge 7 fm}$ at 52 A.MeV), relation
(\ref{eq_tsf}) cannot reproduce the two kinematical peaks of 
the QP emissions. The TSF
has not been applied to these bins.  
The results obtained  at 95 A.MeV for protons at b=5 fm are presented 
in  figure \ref{f:tsf}.
The histograms are the experimental distributions whereas the dotted lines 
are the results of the fit.  The curves are displaced by a factor of 10 in 
order to present all of 
them on the same plot. The agreement 
is particularly good at forward angles.  For backward angles, over 
two orders of magnitude, the
agreement is still good.  
In  figure \ref{f:tsf} (b), the sum  (d$\sigma$/dE) of all 
these curves is shown.  The contribution 
of each source is also presented.   We remark 
that intermediate velocity products extend over a broad energy range
covering partially the region populated by products coming from projectile and 
target deexcitation.
In  figure \ref{f:tsf} (c) the velocity spectrum ( d$\sigma$/dY) of 
each contribution 
shows the predominant role of the  intermediate velocity products. 

\begin{figure}[h]
\begin{center}
\psfig{file=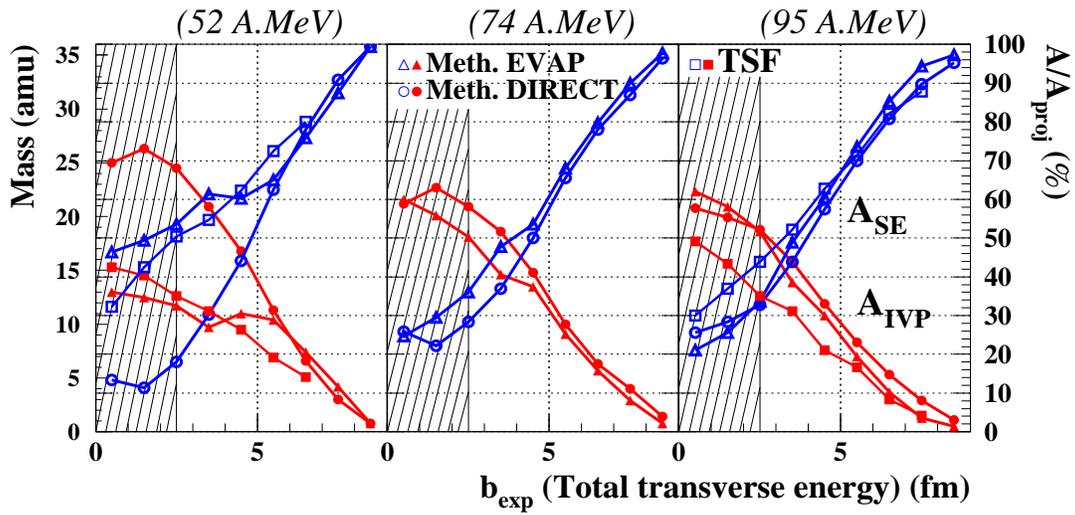,width=15cm}
\caption[]{\label{AMRE}\rm{ Full symbols: total mass of the intermediate 
velocity products emitted above the mid-rapidity ($Y_r > 0.5$) 
as a function of the impact parameter, at 52,
74 and 95 A.MeV, extracted using the three different methods, EVAP
(triangles),  DIRECT (circles)  and TSF (squares). See text for
details. The corresponding masses are given on the left hand scale, whereas
they are given as a percentage of the projectile mass on the right hand scale.
The dashed 
 zones represent 
the impact parameter domain where methods  EVAP and  DIRECT 
are less accurate.}}
\end{center}
\end{figure}

For each impact parameter, the light particle mean multiplicity for
each source is obtained from the deconvoluted spectra like this of figure
\ref{f:tsf}(c).  Since the fragments present a small intermediate velocity
contribution, we will use their mean multiplicities evaluated in a two
source analysis \cite{Dor98}.  We add them to the light particle and neutron
(Mult$_n$ = Mult$_p$) 
mean multiplicities to obtain the total mass of the system.
 In order to compare to the other two methods, 
only the part of the intermediate velocity component above
$\rm{Y/Y_p} \geq$ 0.5 is taken into consideration.
Results are presented in figure \ref{AMRE} (squares).

\subsection{Validity and accuracy.}

The accuracy of methods  EVAP and  DIRECT depends on our capacity 
to distinguish and separate the two components in the rapidity distributions. 
In the most central collisions and for the lowest energy, the damping 
of the two partners is large. So, some of the products emitted from  
 the QP and the QT are localized around mid-rapidity, 
especially the
contribution of QT emission for this slightly asymmetric system. 
These components may lead to overestimate the intermediate velocity 
component.  The mixing of the two components is 
the largest in central collisions 
where the separation in two components  becomes questionable. 
Nevertheless we will present 
the results obtained in the three bins of highest total transverse 
energy but in a dashed zone to keep in mind that the corresponding results are 
less accurate and may be meaningless. 
Finally, since method  EVAP gives a lower limit of the 
number of  intermediate velocity nucleons, and method  DIRECT an upper limit, 
most of the results presented here will be average numbers.

In the most central collisions the mixing of the two components 
does not allow us to determine the QP's rapidity with method  EVAP. 
But we observed that the QP's rapidity decreases regularly from peripheral to 
semi-central collisions ($\rm{b_{exp}}$=3 fm). 
Thus, the QP rapidity was linearly extrapolated from $\rm{b_{exp}}$=3 fm
down to $\rm{b_{exp}}$=0 fm.  This represents a cross section of 280 mb 
or 11\% of the total cross section.

As  mentioned, the TSF method has been  used only on light particles
(p,d,t,$^3\rm{He}$ and alpha). In contrast with methods  EVAP and  DIRECT 
the source velocities are free parameters whose value may depend upon 
the particle  species. 
For a given impact parameter  
 the velocities of the QP and of the intermediate velocity products deduced
from the three sources fit vary within 10$\%$ 
(20$\%$) according to the particle species.
Another difference comes from QT emissions taken into account with the TSF 
method through
the angular distribution.  This lowers the estimated 
 intermediate velocity products contribution compared 
to  EVAP and  DIRECT methods.
In contrast with these methods, TSF 
stays valid for small impact parameters.
 Since the impact parameter sorting could be correlated to the high
kinetic energies of intermediate velocity products, another variable was used
for sorting: the largest charge of the detected products (${\rm Z_{max}}$). The
results are close to those presented here.

In references \cite{Riv96,Bor96},  a pure binary scenario was assumed 
for events containing only light particles (${\rm Z_{max}}$=2). These events
correspond to the most central collisions, and represent a very small part of
the total cross-section. As said above, for these central collisions, the
distinction between the different emission processes becomes questionable.

\section{\label{caract}\textbf{Characterization of intermediate 
velocity products}}

\subsection{Estimations of  masses involved in 
intermediate velocity products and statistical emissions.}
The three methods allow us to determine the  masses involved in 
 intermediate velocity products and in statistical emissions by adding up 
the masses of detected products.
In doing so we take into account the 90$\%$ geometrical 
efficiency and we assume that the number of emitted neutrons is 
equal to the number of detected protons. Finally, we take, for fragments 
with a charge greater than four, the  corresponding mass on the 
$\rm{\beta}$-stability line since for these fragments only the charge 
is measured. 

The  total masses associated respectively to intermediate velocity emission 
and to statistical evaporation,
 above $Y_{nn}= 1/2 Y_{proj}$, obtained with methods 
 EVAP and  DIRECT at 
52, 74 and 95 A.MeV are displayed versus the estimated impact parameter in 
figure \ref{AMRE}. For 52 and 95 A.MeV, the results of the 
TSF method are also 
displayed (full and open squares).

As expected, we note that method  DIRECT gives always a larger estimated 
 intermediate velocity products mass 
than method  EVAP and the TSF  a smaller one. The larger difference between 
 EVAP and  DIRECT 
estimations observed in central collisions at 52 A.MeV is due to the 
lower accuracy of both methods
as explained in section \ref{estimation}.

The  mass of products emitted at intermediate velocity 
increases when the estimated impact parameter 
decreases. The opposite  trend is observed for the mass resulting from 
statistical evaporation. Both 
features are observed for the three methods and are qualitatively 
consistent with theoretical predictions 
carried out on the $\rm{^{40}Ar +~^{27}Al}$ system at 65 A.MeV \cite{Eud97}. 
On the other hand, the decrease with the experimental 
impact parameter of the mass 
associated to a statistical evaporation is different
to what was observed in previous analysis where the mass of the primary 
QP was found nearly constant \cite{Pet95,Ste96,Cus93}. Indeed, 
in these studies all the products emitted above mid-rapidity 
(or center-of-mass rapidity) were used to calculate the QP's rapidity 
(except  hydrogen isotopes which were not used). Due to the 
 presence of the intermediate velocity component 
the QP's rapidities were underestimated
and therefore the obtained values of QP mass were overestimated.

\subsection{Dependence on incident energy of 
 the total mass emitted at intermediate velocity}

\begin{figure}[h]
\begin{center}
\psfig{file=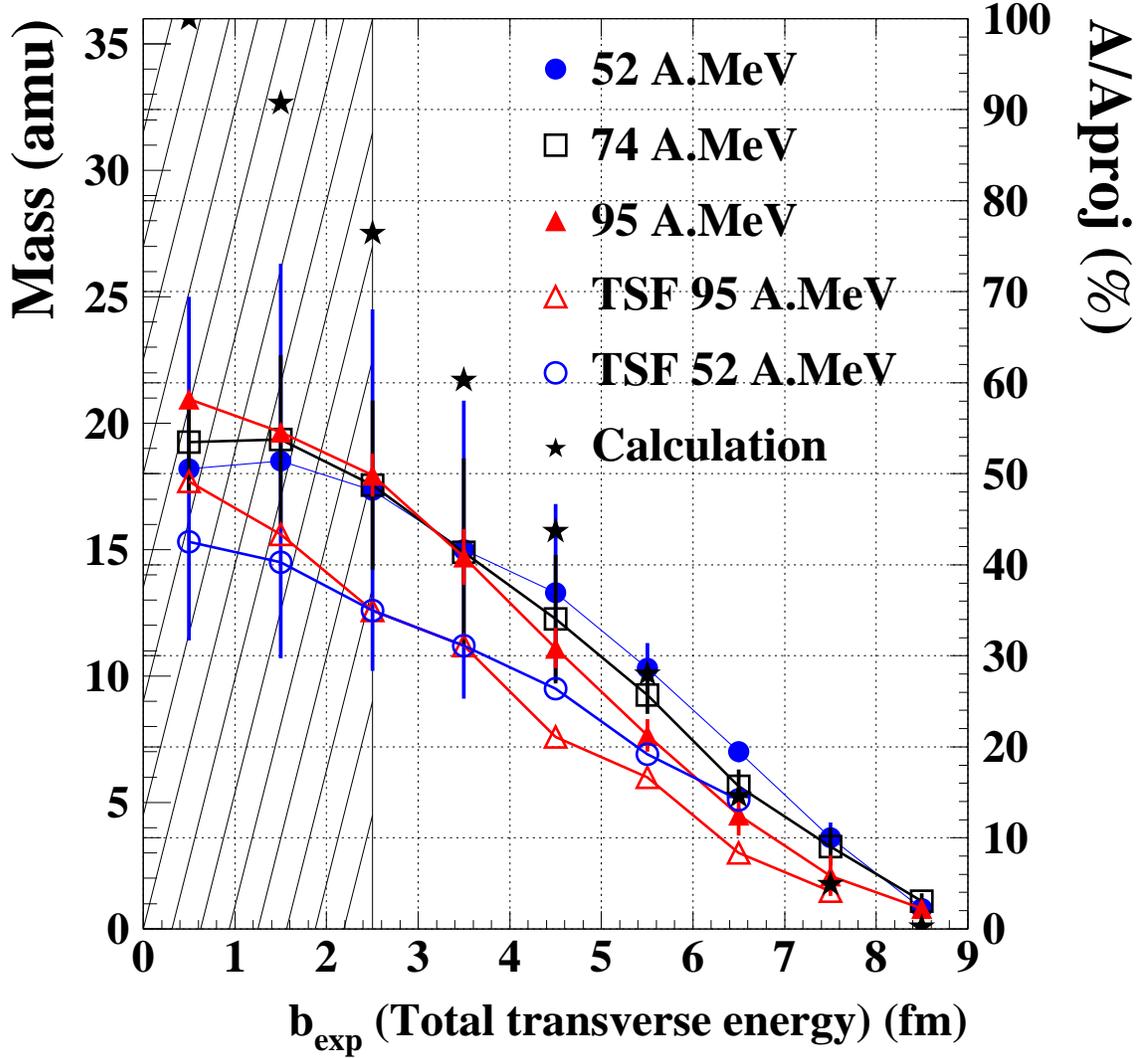,width=16cm}
\caption[]{\label{AMREnrj}\rm{Evolution of  the total mass of the intermediate 
velocity products emitted above the mid-rapidity ($Y_r > 0.5$) 
as a function of 
$b_{exp}$ with  
incident energy. Full circles, open squares and full triangles correspond to 
the mean value
calculated with methods  EVAP and  DIRECT 
(the error bar tips represent inferior
and superior limits given by the two methods), open circles and triangles,  
to the values obtained with TSF method and stars, to a 
participant-spectator prediction for the participant nucleons. 
The left vertical scale represents the absolute mass values, the right
one, to mass  percentage relative to the projectile mass. The dashed 
zone represents the impact parameter domain where 
methods  EVAP and  DIRECT are less accurate.}}
\end{center}
\end{figure}

We also study the influence of the incident energy on the 
 total mass emitted at intermediate velocity. 
In figure \ref{AMREnrj}, we observe that the mean value 
extracted from  method  EVAP and  DIRECT is  not very sensitive on the
incident energy  within the error bars. This result is consistent 
with the constant  shape of triton 
rapidity distributions. 
Nevertheless, owing to the lower 
accuracy of methods  EVAP and  DIRECT at 52 A.MeV, we are more confident 
in the intermediate velocity products total mass constancy from 74 A.MeV 
and above. For the TSF method, the variation of 
the  intermediate velocity products mass
as a function of the impact parameter is the same at 52 and 95 A.MeV. 
 Thus, whatever the method used, it is found that the quantity
of mass emitted at intermediate velocity depends only weakly upon the 
incident energy. 

Finally, to check the influence of the experimental 
impact parameter on the 
 mass emitted at intermediate velocity, we 
 calculated the number of nucleons contained in the overlap region. 
We used an analytical method developed in \cite{Gos77} which determines 
the number of nucleons contained in the volume of interaction of a sphere 
and a cylinder. The geometrical calculations are displayed with black stars 
in figure \ref{AMREnrj}. We note that the geometrical assumption is 
consistent with experimental data for $\rm{b_{exp}}$ values between 4 and 
8 fm. For the most central collisions, the divergences could be due 
to the lower accuracy of the methods used, to the global variable (total 
transverse energy) used to sort the events which is less accurate to sort 
the central ($\rm{b_{exp} \le}$ 3 fm) collisions \cite{Tom97}
or to the existence of a target-like and a projectile-like 
remnants even for those central collisions \cite{Mar97}.

One could argue that the number of intermediate velocity
products is strongly correlated to ${\rm{E^{tot}_{tr}}}$, 
leading to the observation made here. 
The calculations of the correlation factors have shown that
${\rm{E^{tot}_{tr}}}$ is more correlated to the multiplicity 
of the products emitted by
the QP than to the multiplicity of intermediate velocity products. 
As it will
be shown in section \ref{energy}, this can be
explained by a stronger increase of the average transverse kinetic energy 
for the QP emissions than for intermediate velocity emissions. 
Thus the strong 
dependence on the experimental 
impact parameter of the quantity of matter emitted at 
intermediate velocity is not due to a strong correlation 
of this 
mass with the sorting variable ${\rm{E^{tot}_{tr}}}$.
 
In conclusion, the increase  when the estimated impact parameter decreases, 
of the mass emitted at intermediate velocity
is consistent with a geometrical description : the mass involved is 
proportional to the volume of the interaction zone between the colliding 
nuclei except in central collisions ($\rm{b_{exp} <}$ 4 fm).  This 
picture is further supported by the weak dependence
of this emission on the incident energy.

\subsection{Multiplicity of intermediate velocity products}

\begin{figure}[h]
\begin{center}
\psfig{file=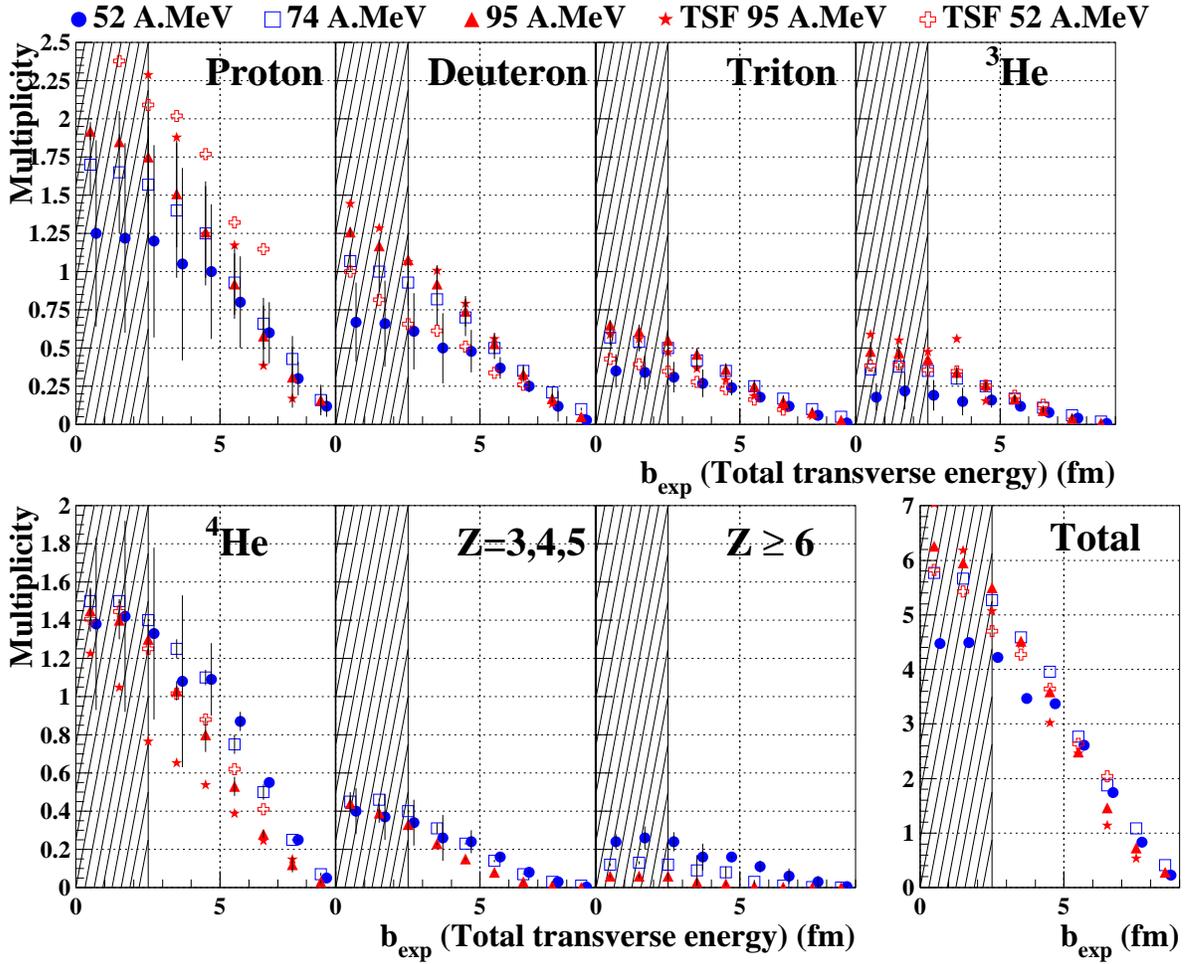,width=16cm}
\caption[]{\label{Mult}\rm{Multiplicities of  intermediate velocity 
products emitted above the mid-rapidity ($Y_r > 0.5$) 
established at 52, 
74 and 95 A.MeV. The mean multiplicities are  obtained by averaging, using
both methods, 
 EVAP and  DIRECT 
and the error bars correspond to the inferior and the superior 
limits. The crosses and the stars correspond to the multiplicities obtained 
with the TSF method at respectively 52 A.MeV and 95 A.MeV. 
The dashed zone represents the impact parameter domain 
where methods  EVAP and  DIRECT are less accurate.}}
\end{center}
\end{figure}

Figure \ref{Mult} shows the absolute multiplicities of the 
intermediate velocity products (mean value obtained from  methods 
EVAP and DIRECT),  and TSF for 52 and 95 A.MeV as a function of 
the estimated impact parameter. A strong dependence is observed for 
the multiplicities of each kind of product. The decrease with $\rm{b_{exp}}$ 
is qualitatively consistent with previous results obtained with the 
system ${^{36}Ar+^{27}Al}$ at the same incident energies 
\cite{Pet 2) 90}. It supports the role played by the geometry of 
the reaction. The dependence on incident energy is  much weaker 
than the experimental impact parameter dependence.  
The multiplicities of alpha particles are systematically lower with
the TSF method than with method  EVAP and  DIRECT.  At small impact 
parameters, the QP source  velocity deduced from the fit of the alpha 
particle spectra is much smaller than the one deduced from the
fits of the other particle spectra. This leads to an important 
evaporative contribution of alpha particles and a lower
intermediate velocity component  

 The  intermediate velocity component is  mainly made up of 
light  charged particles (85-95\%  in number)
whatever the impact parameter and the incident energy. Most 
of these particles are protons and alpha particles  which represent 
respectively 30-40\% and 20-30\%  in number and 
10-20\% and 30-40\%  in mass of the intermediate velocity products. 

\section{\label{energy}
\textbf{Energy dissipated in the intermediate velocity 
area}}

The temperature measurements are commonly used to characterize energetic 
properties of nuclei in which thermal (and chemical) equilibrium is 
achieved \cite{Mor94}. Recently, the caloric curve shown by the 
ALADIN group \cite{Poc95} has stimulated many theoretical and experimental 
studies about the accuracy of the different thermometers used
\cite{Ma97,Nat95,Mor96,Haug96,Kwia98,Siw98,Gul97}. 
In this section, we will present temperature measurements obtained for
intermediate velocity products. Since there 
is no evidence around mid-rapidity for the existence of a  
source in which thermal equilibrium is attained, we will just use the 
measured thermometer values as indications of the energy
dissipated in 
the intermediate velocity 
area. Thus, only the relative
variations of the values of these thermometers will be discussed.

\subsection{Definition of parameters used.}

We use two kinds of methods. Firstly, we determine a parameter
obtained from the production yields ($\mathcal{P}$) of several pairs of light 
isotopes differing by one neutron (Double Ratio Parameter labeled DRP), 
according to \cite{Alb85}:

\begin{equation}
DRP=\Delta B/\log \left( a.\frac{{\mathcal P}(Z_n,A_n)/{\mathcal P}(Z_n,A_n+1)}
{{\mathcal P}(Z_d,A_d)
/{\mathcal P}(Z_d,A_d+1)} \right) 
\label{eq_DRP}
\end{equation}

where $n$ ($d$) stands for the pair of isotopes with the smallest (largest)
binding energy difference, and appears at the numerator (denominator)
on the right  hand side. $\Delta B$ is the difference of binding energy 
differences and $a$ depends on the spins and on the masses 
of the populated states. For 
more details see \cite{Poc95,Alb85}. If $DRP$ is calculated from the 
emitted products of an equilibrated (thermally 
and chemically) source, 
then $DRP$ is a measure of the temperature of this source.
Nonetheless, the secondary decay of excited emitted fragments 
(side-feeding effect) lowers considerably the measured temperature with 
respect to the initial temperature of the source \cite{Siw98,Xi99}. 

Secondly, we determine the slope of kinetic energy distributions of 
 intermediate velocity products. If the kinetic energy 
distributions are established in the frame of an equilibrated source, 
then their inverse slope parameter $\mathcal{S}$ is
a measure of the temperature of this source. The inverse slope parameters
were obtained via the usual fits with Maxwell-Boltzmann distributions.
The surface emission formula \cite{Gol78} has been used:

\begin{equation}
P(Ec)=\alpha \frac{(Ec-B)}{{\mathcal S}^2} exp(- \frac{Ec-B}{{\mathcal S}}) 
\label{eq_TPente} 
\end{equation}
where $\rm{\alpha}$ is a normalization factor, $Ec$ is the kinetic energy of 
the emitted product in the source frame, $B$ the  Coulomb barrier between 
the source and the emitted product and $\mathcal{S}$ the temperature of 
the source. The obtained temperature is averaged over the de-excitation 
chain of the source and affected by successive recoil effects. 
The apparent temperatures are therefore lower than the initial 
ones \cite{Siw98}. The use of a volume emission formula, as for the
three source fit method exposed in section \ref{caract}, leads to
very close values of the slope parameter.

We remind  the reader that there is no evidence for the existence 
of an equilibrated source in the mid-rapidity region.  Thus, to 
avoid any confusion, we will not associate a thermodynamical temperature 
to the parameters $DRP$ and $\mathcal{S}$. 
We will use them as indicators of 
the energy dissipated in the intermediate velocity area.

\subsection{Dependence on the violence of the collision}
\begin{figure}[h]
\begin{center}
\psfig{file=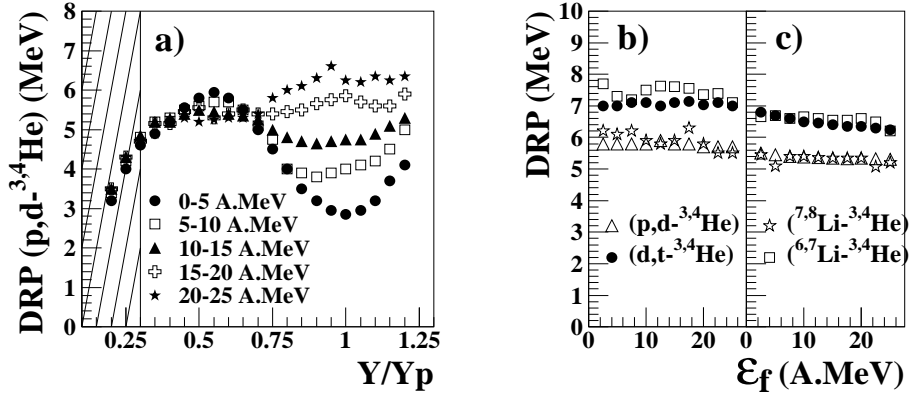,width=14cm}
\caption[]{\label{Tempratio}\rm{Left panel: double ratio parameter 
($\rm{p,d-^{3,4}He}$) versus the relative rapidity. It has been calculated 
for five bins 5 MeV wide in ${\mathcal E}_{\rm f}$. The dashed zone corresponds 
to the rapidity domain where the identification of emitted products is 
affected by thresholds. Right panel: four isotope double ratio parameters 
versus ${\mathcal E}_{\rm f}$ for two  intermediate velocity
particle selections: 
 an angular cut (b), a rapidity  cut (0.35 < $Y_r$ < 0.65) (c). 
This figure was established at 95 A.MeV.}}
\end{center}
\end{figure}

In the first stage, we  determined the average double ratio parameters 
($\rm{p,d-^{3,4}He}$) over the whole range of relative rapidity,
except in the low rapidity region where thresholds affect  
the identification of light isotopes. These  ratios were determined 
at 95 A.MeV for five bins in ${\mathcal E}_{\rm f}$ distribution  
(see equation \ref{eq_Ef} in section \ref{sorting_b}). 
The width of each bin is equal to 5 MeV. The results are displayed 
in figure \ref{Tempratio} a).

The double ratio parameters measured around the QP rapidity
increase with the violence of the collision. This fact is known 
and has been observed with different double ratio and slope 
parameters on the same system for the caloric curve study  
\cite{Ma97}. 
On the contrary, we note that the double ratio parameters  at mid-rapidity 
are nearly constant whatever the violence of the collision. 
Thus it appears that the evolution with ${\mathcal E}_{\rm f}$ of the 
double ratio parameters for intermediate velocity products 
is different  from the evolution of double ratio parameters of
statistical emission. 

 In order to confirm the near-constancy of the double ratio parameters 
for intermediate velocity products, it is
necessary to isolate them to minimize the contribution from QP and QT 
statistical emission. For this purpose, two different
cuts were applied : i) an angular cut where only products emitted at 
90$^\circ  \pm $ 15$^\circ$ in the mid-rapidity frame
are considered; ii) a rapidity cut where only products emitted in the 
rapidity region 0.35 $\leq$ $\rm{Y/Y_p}$ $\leq$ 0.65 are
kept. The four double ratio parameters from the following 
combination ($\rm{p,d-^{3,4}He}$), 
($\rm{d,t-^{3,4}He}$), ($\rm{^{7,8}Li-^{3,4}He}$) 
and ($\rm{^{6,7}Li-^{3,4}He}$) are presented in fig. 10b) and 10c) 
using respectively the angular cut and the rapidity cut.
 
We note in figures \ref{Tempratio} b) and \ref{Tempratio} c) that with 
both selections the four double ratio parameters are insensitive to 
${\mathcal E}_{\rm f}$. This feature  is observed whatever the selection 
used. Since there is no major difference between both selections, we will 
only use the angular cut in the remaining part of the article.

To check the independence on ${\mathcal E}_{\rm f}$, of the energetic properties
of the  intermediate velocity products, we use another ``thermometer'':
the slope parameters of kinetic energy spectra of  intermediate velocity 
products in the  mid-rapidity frame. We  obtained 
them for three incident energies (52, 74 and 
95 A.MeV) and for five light particles: p, d, t, $\rm{^3He}$ and $\rm{^4He}$. 
The results are displayed in figure \ref{Tempslop}. 
 A weak dependence on ${\mathcal E}_{\rm f}$ is observed. 

Finally, we have checked that this observation is independent of the global 
variable used to estimate the violence of the collision. Owing to the 
strong correlation between the total transverse energy 
and the kinetic 
energy of emitted products, it was not meaningful to measure 
the slope parameters of light particles with this sorting. That is
why we determined only double ratio parameters. Figure \ref{Temprationrj} 
shows four double ratio parameters versus the estimated impact parameter. 
We note that the same weak dependence is observed.  With the 
TSF method, the pd/$^3$He$^4$He and pt/$^3$He$^4$He for the intermediate 
velocity products are also  roughly constant according to $\rm{b_{exp}}$. 
These values are slightly different from the values obtained with the two 
other methods.

One can notice that the same system has been used for the study
of the caloric curve in \cite{Ma97}. But the temperature values obtained 
from both analysis can not be directly compared. In \cite{Ma97} the 
temperature values have been extracted from the QP desexcitation whereas 
in this article the $DRP$ and $\mathcal{S}$ values come from the intermediate 
velocity 
region. In addition, the data in \cite{Ma97} are consistent with the 
decay of an equilibrated source. For the intermediate velocity source, 
we do not have any proof that an equilibrium has been reached. 
Nonetheless, in the 
most central collisions of \cite{Ma97}, one notes that the temperature 
values obtained are close to the DRP and S values of the intermediate 
velocity region. As already mentioned, for the central collisions it 
is hard to distinguish between the QP or QT emissions 
and the intermediate 
velocity emissions 
with our global variable. Therefore a possible mixing 
may explain this similarity.

In conclusion,  the energetic properties of the intermediate velocity products 
are nearly independent of the violence of the collision whatever the 
thermometer and the sorting variable used. These results indicate that 
the available energy per nucleon for the production of light products 
at  intermediate velocity is nearly constant over the whole range of 
impact parameters, i.e. it is independent of the collision geometry. 

\subsection{Dependence on the incident energy}

In order to understand the origin of  intermediate velocity products, 
it is also useful to study the dependence on the incident energy 
of their energetic properties. Indeed any dynamical process 
should give kinetic energies directly correlated to the incident 
energy. 

\begin{figure}[h]
\begin{center}
\psfig{file=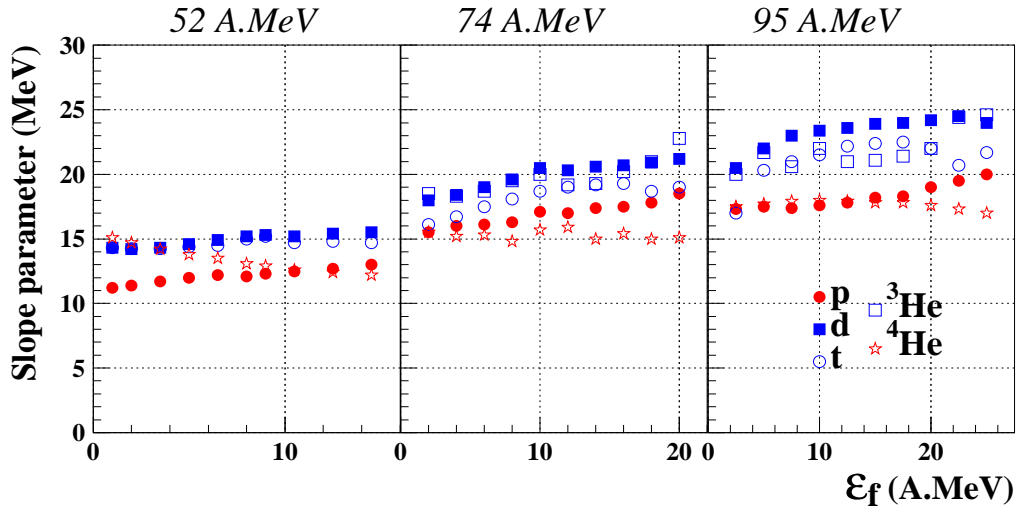,width=14cm}
\caption[]{\label{Tempslop}\rm{Slope parameters calculated from the kinetic
energy spectra of the light particles at 52, 74 and 95 A.MeV versus 
${\mathcal E}_{\rm f}$.  Intermediate velocity products are selected 
 with the angular cut defined in text.}}
\end{center}
\end{figure}

\begin{figure}[h]
\begin{center}
\psfig{file=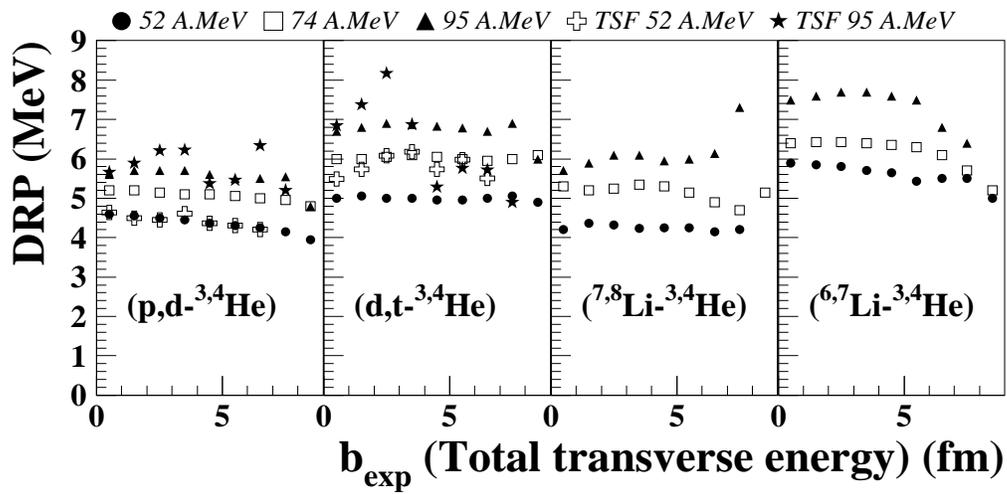,width=14cm}
\caption[]{\label{Temprationrj}\rm{Four double ratio parameters calculated at 
52, 74 and 95 A.MeV versus the estimated impact parameter. 
 Intermediate velocity products
are selected  with the angular cut defined in text. 
 The stars and the crosses
correspond to the values obtained with the three source fit method.}}
\end{center}
\end{figure}

\begin{figure}[h]
\begin{center}
\psfig{file=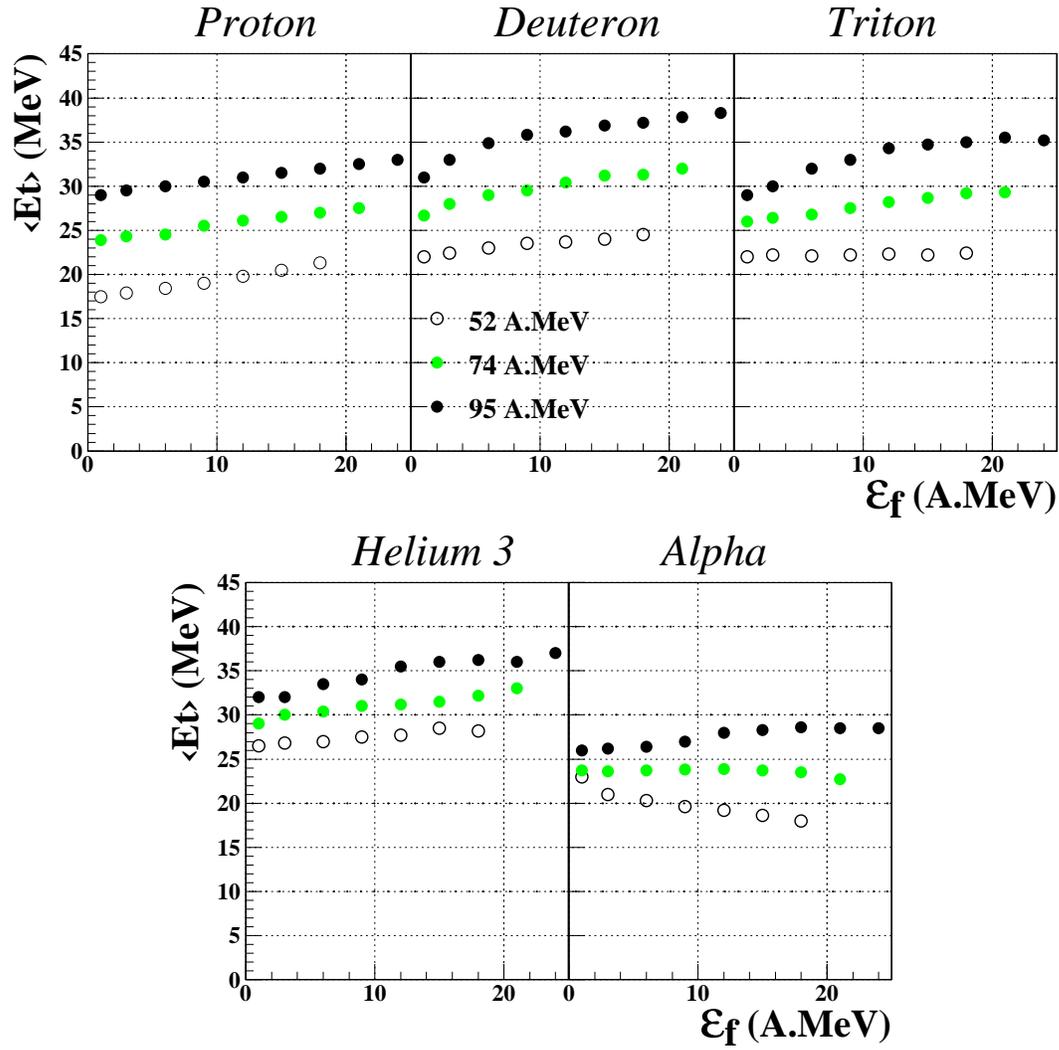,width=14cm}
\caption[]{\label{Etnrj}\rm{Mean transverse energy of light particles
(p, d, t, $\rm{^3He}$ and $\rm{^4He}$) calculated at 52, 74 and 95 A.MeV 
versus ${\mathcal E}_{\rm f}$.  Intermediate velocity products are selected 
in the angular range  defined in text.}}
\end{center}
\end{figure}

Figures \ref{Tempslop} and \ref{Temprationrj} show that slope 
parameters and double ratio parameters increase with the incident 
energy. The same trend has been observed in \cite{Che87}. In order 
to confirm this behaviour, we calculate the mean transverse energy 
of light particles versus ${\mathcal E}_{\rm f}$ at 52, 74 and 95 A.MeV.
Intermediate velocity products are selected in the previously 
defined  angular cut. The results are displayed in figure \ref{Etnrj}. 
We note that, except in peripheral collisions, the mean transverse energy 
of light particles increases proportionally to the incident energy. 
One can also  notice that the mean transverse energy values calculated 
in figure \ref{Etnrj} are higher than those observed in figures 
\ref{rapet1} and \ref{rapet2}. This is due to the angular selection 
which avoids products with the lowest transverse energy and therefore 
increases the mean transverse energy values.

In the case of a thermal emission at intermediate velocity, one
should expect roughly the same mean kinetic energy value for all the
emitted light charged particles ($\rm{<E>=~B_c+ 2 \times T}$ where
$\rm{B_c}$ 
is the coulomb barrier and T the temperature of the source). An overall
agreement can be observed in figure
\ref{Etnrj} for the deutons, tritons and Helium 3 but the alpha particles
are systematically lower than the others. The protons are somewhere between
the alpha particles and the d,t and $\rm{^3He}$. The same general
behaviour is observed in figure \ref{Tempslop}. A sole thermal
emission is thus not consistent with the measured kinetic energies
of intermediate velocity products.

In the case of a prompt emission, one expects the following mean
kinetic energy at mid-rapidity:

\begin{equation}
 \rm{<E>= \frac{1}{2} m \left( \frac{V_{proj}}{2} \right)^2~+ <E_{Fermi}>}
\end{equation}

where $m$ is the mass of the scattering particles and $\rm{<E_{Fermi}>}$
is the mean kinetic energy (= 12 MeV \cite{Fuc94}) of the Fermi motion.
This leads to a mean kinetic energy at mid-rapidity equal to 25 MeV, 30
MeV and 36 MeV at respectively 52, 74 and 95 MeV/A.
One notes that the measured mean kinetic energy of protons are lower
than the values obtained with a pure pompt emission. Unlikely, the mean
kinetic energy of d, t and $\rm{^3He}$ are more consistent. However, there is
again no agreement between all the light charged particles. This latter
observation confirms that intermediate velocity products come from
different processes.

In summary, the energy dissipated in the intermediate 
velocity area  
is weakly dependent on the violence of the collision and 
is mainly governed
by the incident energy. These properties support a dynamical origin
for the intermediate velocity emissions or a participant-spectator scenario.

\section{\label{summar}\textbf{Summary}}
The present work is an attempt to characterize the  products emitted 
at intermediate velocity in the 
$\rm{^{36}Ar+^{58}Ni}$ reactions between 52 and 95 A.MeV using 
the INDRA detector.

The analysis of kinematical properties of light particles and 
intermediate mass fragments in both directions (perpendicular and 
parallel to the beam axis)  suggests a particular type of emissions
located around mid-rapidity. The significant differences observed 
between the experimental data and the calculations performed with 
a statistical code assuming two thermalised sources have clearly 
pointed out the existence of the
 intermediate velocity products 
(which could be due to several processes). Most of 
emitted deuterons and tritons come from those emissions.

Three methods have been  developed to distinguish and separate two
components: the  intermediate velocity products 
which can include several
emission processes, 
and the statistical
emission from the fully equilibrated QP decay. 
We  have determined the amount of
matter associated to  intermediate velocity products:  their total 
mass is directly correlated to the
impact parameter and  depends only weakly on the incident energy.
The  total mass of the intermediate velocity products represents roughly 
30-40\% of the initial projectile mass 
in mid-central collisions. This result indicates that any study of
the completely equilibrated QP decay has to be very carefully done.
 Finally,  the total mass of the intermediate velocity products is 
mainly governed 
by the geometry of the collision from 52 up to 95 A.MeV. 
These properties suggest that the overlap zone 
of the two colliding nuclei is the origin of  the intermediate 
velocity products.

Finally, we  have examined the energy dissipated by the 
light particles and 
intermediate mass fragments in  the  intermediate velocity zone. 
The energy dissipated by the 
intermediate velocity products is weakly dependent on impact 
parameter but is directly correlated to the incident energy. These 
properties are consistent  with dynamical origins for 
 intermediate velocity products. 

 The three scenarii described in the introduction are consistent with 
the observed properties:  i) prompt emission, i.e. products are 
emitted by direct collisions in the first steps of the collision; 
 ii) formation and rupture 
of a neck of matter between the two colliding nuclei;
 iii) a participant-spectator model. 
These three  scenarii are idealistic. 
Obviously, in the same event, the prompt
emission can  precede the formation of a participant zone or a 
neck of matter. It is also possible that other exotic processes take place. 
Dynamical calculations could be useful to check these different assumptions.

\newpage

\end{document}